\documentclass[aps,prl,reprint]{revtex4-2}
\usepackage{blindtext}

\usepackage[utf8]{inputenc}
\usepackage{graphicx,fancybox,float,comment}

\usepackage{units}
\usepackage{subcaption}
\usepackage{multirow,array,arydshln}
\usepackage[dvipsnames]{xcolor}
\usepackage{amsmath,amssymb,amsfonts}
\usepackage{yfonts}
\usepackage{tensor}
\usepackage{paralist}
\usepackage{braket}
\usepackage{tensor,mathrsfs}
\usepackage[normalem]{ulem}

\usepackage{hyperref}
\usepackage{cleveref}

\newcommand{\om}{\Omega}
\newcommand{\qn}{\mathfrak{q}}
\newcommand{\wn}{\mathfrak{w}}

\begin{document}
\title{Convergence of hydrodynamics in rapidly spinning strongly coupled plasma}

\newcommand{\uaAff}{\affiliation{
Department of Physics and Astronomy, 
University of Alabama, 514 University Boulevard, Tuscaloosa, AL 35487, USA}}
\newcommand{\illAff}{\affiliation{
Illinois Center for Advanced Studies of the Universe, 
Department of Physics, University of Illinois at Urbana-Champaign, Urbana, IL 61801, USA}}

\author{Casey Cartwright} \email{cccartwright@crimson.ua.edu} \uaAff
\author{Markus Garbiso Amano} \email{magarbiso@crimson.ua.edu} \uaAff
\author{Matthias Kaminski} \email{mski@ua.edu} \uaAff
\author{Jorge Noronha} \email{jn0508@illinois.edu} \illAff
\author{Enrico Speranza} \email{espera@illinois.edu} \illAff

\date{\today}

\begin{abstract}
We compute the radius of convergence of the linearized relativistic hydrodynamic expansion around a non-trivially rotating strongly coupled $\mathcal{N}=4$ Super-Yang-Mills plasma. Our results show that the validity of hydrodynamics is sustained and can even get enhanced in a highly vortical quark-gluon plasma, such as the one produced in heavy-ion collisions. The hydrodynamic dispersion relations are computed using a rotating background that is an analytic solution of the ideal hydrodynamic equations of motion with non-vanishing angular momentum and large vorticity gradients, giving rise to a particular boost symmetry. Analytic equations for the transport coefficients of the rotating plasma as a function of their values in a plasma at rest are given.
\end{abstract}

\maketitle


\emph{Introduction.}---Relativistic hydrodynamics is a powerful tool to describe the late time, long wavelength behavior of the strongly coupled quark-gluon plasma (QGP) formed in ultrarelativistic heavy-ion collisions \cite{Heinz:2013th,Romatschke:2017ejr,Florkowski:2017olj}. Recent experimental results show that certain hadrons (e.g.~Lambda-hyperons) emitted in noncentral collisions exhibit nonzero spin polarization, which indicates that the QGP is the most vortical fluid observed to date~\cite{STAR:2017ckg}. Despite some early successes \cite{Becattini:2020ngo}, a number of questions concerning the vorticity of the QGP still remain, see e.g. \cite{Becattini:2020ngo,Florkowski:2017ruc,Florkowski:2017dyn,Montenegro:2017rbu,Becattini:2018duy,Florkowski:2018fap,Hattori:2019lfp,Bhadury:2020puc,Montenegro:2020paq,Weickgenannt:2020aaf,Speranza:2020ilk,Garbiso:2020puw,Fukushima:2020ucl,Li:2020eon,Gallegos:2021bzp,Speranza:2021bxf,Hongo:2021ona}. One crucial aspect yet to be understood is how the large angular momentum present at the early stages of the collision influences the emergence of hydrodynamics in the presence of vorticity. A fundamental related question is then: what is the range of validity of hydrodynamics in a rapidly rotating strongly coupled QGP? This is currently beyond the reach of standard first-principles approaches \cite{Meyer:2011gj}, as it requires knowledge about the real-time properties of QCD in the strongly coupled regime. 

In this work we answer this question for a well-established ``toy-model" of the rotating QGP, i.e., the strongly coupled $\mathcal{N}=4$ Super-Yang-Mills (SYM) plasma at nonzero temperature and angular momentum, using the gauge/gravity correspondence (holography)~\cite{Maldacena:1997re}. This model consists of a rotating plasma of quarks and gluons (in the adjoint representation) of $\mathcal{N}=4$ SYM theory with gauge symmetry $SU(N_c)$ in the limit of an infinite number of colors, $N_c\to\infty$, and a large ('t Hooft) coupling constant $\lambda$. By investigating the critical points of spectral curves as developed in Refs.~\cite{Grozdanov:2019uhi,Grozdanov:2019kge}, we show that the radius of convergence of the hydrodynamic series, which quantifies the range of validity of the (linearized) hydrodynamic gradient expansion, remains finite at nonzero angular momentum. The radius of convergence can even increase for a sufficiently rapidly rotating plasma. Furthermore, we demonstrate that the transport coefficients of the rotating plasma can be explicitly obtained from their values in the plasma at rest.

\emph{Holographic model.}---
In order to determine the values of hydrodynamic transport coefficients, we compute hydrodynamic modes and correlation functions of $\mathcal{N}=4$ SYM plasma with the holographically dual Einstein gravity $S=1/(16\pi G_5) \int d^5x \sqrt{-g}(\mathcal{R}-2\Lambda)$, where $\mathcal{R}$ is the Ricci scalar, $G_5$ is the five-dimensional gravitational Newton constant, and $\Lambda$ the cosmological constant ~\cite{Policastro:2002se,Policastro:2002tn}. 
A rotating plasma is holographically dual to the rotating black hole metric which is asymptotic to AdS$_5$ spacetime~\cite{Hawking:1998kw,Hawking:1999dp}. In general, such a black hole is parameterized by two angular momenta, $a$ and $b$, and it has the structure  $\mathbb{R}_t\times \mathbb{R}_r \times S^3$ with a time coordinate  {$t\in(-\infty,\infty)$}, the radial AdS coordinate {$r\in (0,\infty)$}, $r_H$ being the event horizon radius, and the spatial 3-sphere $S^3$ parameterized by angles $(\theta,\phi,\psi)$. We set $b=a$~\footnote{Note that $a$ here is defined with a relative minus sign compared to the $a$ used in \cite{Garbiso:2019uqb}.} and use a coordinate transformation (diffeomorphism)~\cite{Garbiso:2020puw} yielding the rotating metric in the form~\cite{Murata:2007gv,Murata:2008xr,Murata:2008yx} 
\begin{eqnarray}\label{eq:rotatingMetricSigmas}
   ds^2 &=& -\left( 1 + \frac{r^2}{L^2} \right) {dt}^2 + \frac{{dr}^2}{G(r)} + \frac{r^2}{4} \left( (\sigma^1)^2 + (\sigma^2)^2  \right. \nonumber \\
   && \left . + (\sigma^3)^2 \right)+  \frac{2 \mu}{r^2} \left(dt + \frac{a}{2} \sigma^3 \right)^2\, \nonumber \\
    G(r) &=&  1+\frac{r^2}{L^2}-\frac{2 \mu (1-a^2/L^2)}{r^2}+\frac{2 \mu a^2}{r^4}\, , \nonumber \\ 
    \mu &=& \frac{r_+^4 \left(L^2+r_+^2\right)}{2 L^2 r_+^2-2 a^2 \left(L^2+r_+^2\right)}  \, ,
\end{eqnarray}
with the radius of AdS, $L$, the AdS radial coordinate $r$, the horizon radius $r_+$, the one-forms $\sigma^1, \sigma^2, \sigma^3$, which each are known covectors that depend on $(\theta,\phi,\psi)$~\cite{Garbiso:2020puw}. Throughout this work we use natural units, $c=\hbar= k_B=1$, and  $G_5=L^3 \pi /(2 N_c^2)$.   

\emph{Rotating equilibrium state.}--- 
In $\mathcal{N}=4$ SYM theory, the rotating gravitational metric solution in the large black hole limit 
\begin{equation}\label{eq:largeBHLimitR}
r_+\to \alpha r_+\, , \quad 
r\to \alpha r\, , \quad \alpha\to\infty \, , 
\end{equation}
corresponds to a rotating conformal fluid solution of the ideal hydrodynamic equations of motion $\nabla_\mu T^{\mu\nu}=0$, with total angular momentum of magnitude $J = a\pi\mu/G_5$ \cite{Papadimitriou:2005ii}. 
In Milne coordinates $(\tau,x_1,x_2,\xi)$ of flat Minkowski space~\cite{Romatschke:2017ejr}, 
this flow can be written as 
\begin{align}
    u^\tau &=\lambda\left[ \cosh \xi  \left(L^2+\tau ^2+x_\perp^2\right) \right. \nonumber\\
    &\left. +2 \om  (L x_1 \sinh \xi +\tau  x_2) \right] \nonumber\\
    u^1 &= \lambda\left[2 (L \tau  \om  \sinh \xi +\tau  x_1 \cosh \xi +x_1 x_2 \om )\right] 
    \, ,\nonumber\\
    u^2&=\lambda\left[\om  \left(L^2+\tau ^2-x_1^2+x_2^2\right)+2 \tau  x_2 \cosh \xi \right] \, , \label{eq:fluid_velocity}\\
    u^\xi&=-\tau^{-1}\lambda \left[-\sinh \xi  \left(L^2-\tau ^2+x_\perp^2\right)-2 L x_1 \om  \cosh \xi \right] \, ,  \nonumber\\
   \epsilon&= (16 L^8 \Theta^4)\left(1-\om ^2\right)^{-2}\times \nonumber \\
   &\left(2 L^2 \tau ^2 \cosh 2 \xi +\left(L^2+x_{\perp}^2\right)^2+\tau ^4-2 \tau ^2 x_{\perp}^2\right)^{-2} \, , \label{eq:energy_density}\\
   \lambda&=\left( \frac{\epsilon}{16 L^8\Theta^4}\right)^{1/4}, \quad \Theta=\left(\frac{3(1-\Omega^2)\mu }{8\pi G_5 L^3}\right)^{1/4} \, ,\label{eq:parameters}
\end{align}
where $\om=a/L$, $x_\perp^2=x_1^2+x_2^2$, $\tau=\sqrt{t^2-x_3^2}$, $\xi = \mathrm{arctanh}(x_3/t)$, and we recall that $\mu$ has dimensions $L^2$ and $\Theta$ carries units of energy because $G_5\sim \ell_P^3$, with the Planck length, $\ell_P$. 
This is a specific version of the fluid flow found in~\cite{Bantilan:2018vjv}.\footnote{A distinct holographic dual of a rotating fluid has also been considered by boosting a static black brane~\cite{Chen:2020ath}.}
\begin{figure*}[htbp]
    \begin{center}
    \includegraphics[width=0.25\textwidth]{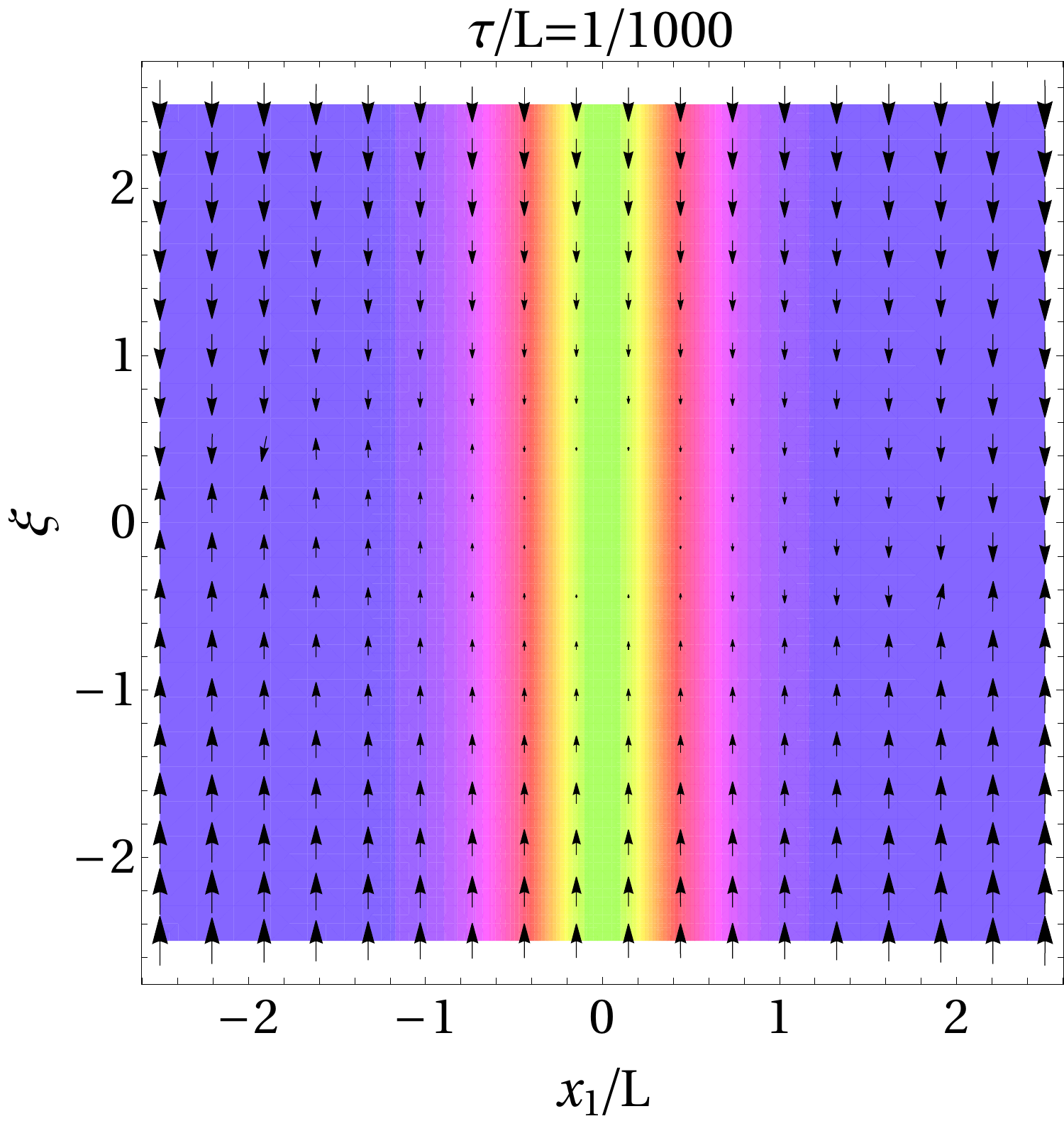} \hspace{0.5cm}\includegraphics[width=0.25\textwidth]{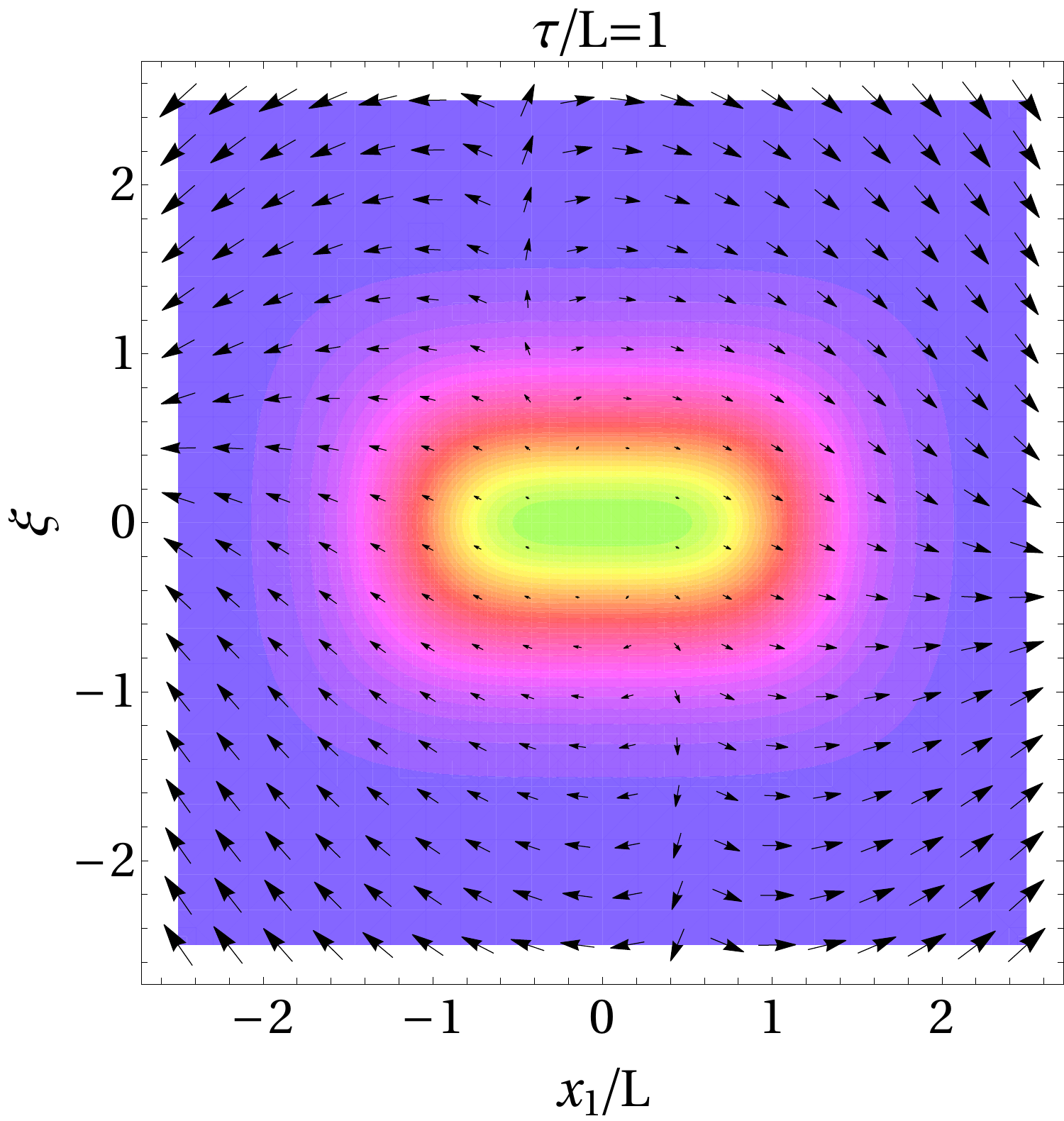}\hspace{0.5cm}
    \includegraphics[width=0.3425\textwidth]{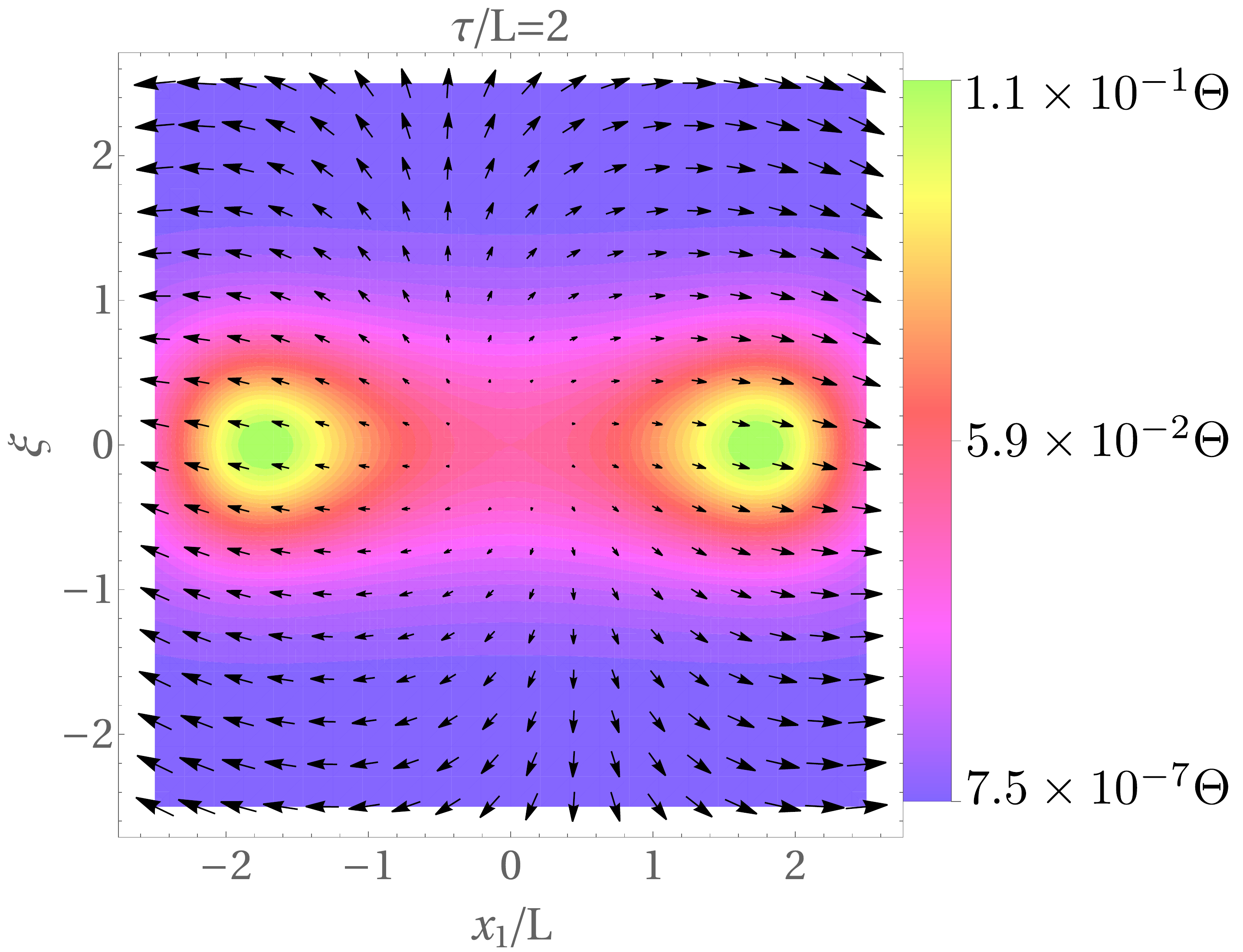} 
    \end{center}
   \caption{\label{fig:rotational_fluid}
    Flow velocity vectors for different times and $\om=1/2$ in the $x_1$-$\xi$ plane (with $x_2=0$). The shading corresponds to level sets of the energy density (where $\Theta$ is the overall energy scale introduced in (\ref{eq:parameters})). We have rescaled the energy density by a factor of approximately 250 in the first plot, and 15 in the second plot for the ease of visualization.
     }
\end{figure*}

This flow encodes a non-trivial rotation of the fluid and is obtained from the eigenvalue equation for the energy momentum tensor $T^{\mu}_\nu u^\nu = -\epsilon u^\mu$ of $\mathcal{N}=4$ SYM theory computed in the large black hole limit~\eqref{eq:largeBHLimitR} after stereographic projection from $\mathbb{R}_t\times S^3$ to $\mathbb{R}^{1,3}$.
Initially, the energy density (coloring) is uniformly distributed in rapidity, see left plot in Fig.~\ref{fig:rotational_fluid}. 
With time progressing, at $\tau/L=1$, 
see center plot in Fig.~\ref{fig:rotational_fluid}, the energy is now concentrated at mid-rapidity and the fluid velocity spirals outward. Finally at $\tau/L$, see right plot of Fig.~\ref{fig:rotational_fluid}, the energy density splits into two outgoing pieces at mid-rapidity. One can compute explicitly that the vorticity $\omega_{\text{vor}}^{i}=\frac{1}{2}\epsilon^{ijkl}u_j \nabla_{[k}u_{l]}$ is non-vanishing for all components 
and can attain large gradients  
throughout the evolution described above.
With these properties, this inhomogeneous, time-dependent flow profile resembles the flow expected for a quark-gluon plasma created in a heavy-ion collision~\cite{STAR:2017ckg}. Note that \eqref{eq:fluid_velocity} contains also non-trivial rotations in the transverse plane, the $x_1-x_2$ plane.

\emph{Hydrodynamic fluctuations.---}
In this work, we are interested in hydrodynamic modes, i.e.,~modes which have a vanishing frequency at vanishing momentum. 
These hydrodynamic modes on the gravity side correspond to a subset of the quasinormal modes (QNMs) of metric fluctuations around the rotating background~\eqref{eq:rotatingMetricSigmas}. 
Consider the metric fluctuations around~\eqref{eq:rotatingMetricSigmas}, which satisfy the linearized Einstein equations~\cite{Wald:1984rg}
\begin{equation}\label{eq:pertgenericeom}
    \begin{aligned}
        \dot{R}_{\mu\nu} = \frac{2\Lambda}{D-2}h_{\mu\nu}\, ,
    \end{aligned}
\end{equation}
where $$\dot{R}_{\mu\nu} = -\frac{1}{2}\nabla_\mu \nabla_\nu h-\frac{1}{2}\nabla^\lambda \nabla_\lambda h_{\mu\nu}+\nabla^\lambda \nabla_{(\mu}h_{\nu)\lambda}\,,$$ and $h=h^{~~\mu}_{\mu}=h_{\nu \mu} g^{\mu \nu}$.  The diffeomorphism-covariant derivatives are defined with respect to the background metric~\eqref{eq:rotatingMetricSigmas}. 
The rotating black hole metric~\eqref{eq:rotatingMetricSigmas} has a spatial $SU(2)\times U(1)$ symmetry  such that Wigner-D functions, $D^\mathcal{J}_\mathcal{KM}(\theta,\phi,\psi)$, form an orthonormal basis on the $S^3$. This is similar to the more familiar spherical harmonics on $S^2$.  
Hence, we choose to expand all metric fluctuations in terms of Wigner-D functions. 

The fluctuation equations~\eqref{eq:pertgenericeom} can be separated into three sectors which decouple because of their different Wigner-charges: 
the {\it tensor sector} contains $h_{++}$ with $\mathcal{K'}=\mathcal{J}+2$, 
the {\it momentum diffusion sector} contains $h_{+t},\, h_{+3}$ with $\mathcal{K'}=\mathcal{J}+1$ ($h_{++}$ will decouple in the large black hole limit we will consider), while 
the {\it sound sector} contains $h_{+-},\, h_{tt},\, h_{t3},\, h_{33}$ with $\mathcal{K'}=\mathcal{J}$ ($h_{++},\, h_{+3},\, h_{+t}$ will decouple in the large black hole limit), in radial gauge $h_{\mu r}\equiv 0$. 
This yields the following forms for the tensor, vector, and scalar fluctuations, respectively~\footnote{Note that $\mathcal{K}'$ is the $\mathcal{K}$-charge of $h_{\mu\nu}$. 
This is in general distinct from the value $\mathcal{K}$ in the corresponding Wigner-D function, because each $\sigma^\pm$ carries a $\mathcal{K}$-charge of $\pm 1$.
} 
\begin{equation}\label{eq:pertsimplyglobaltensor}
    h^T_{\mu\nu} \equiv e^{-i\omega \tau} r^2 h_{++}(r) \sigma^+_{\mu} \sigma^+_{\nu} D_{\mathcal{J} \mathcal{M}}^\mathcal{J} 
    \, ,
\end{equation}
\begin{equation}\label{eq:pertsimplyglobalvector}
    \begin{aligned}
        h^V_{\mu\nu} \equiv & e^{-i\omega \tau} r^2  (h_{++}(r) \sigma^+_{\mu} \sigma^+_{\nu} D_{(\mathcal{J}-1)\mathcal{M}}^\mathcal{J} + \\
        & 2 (h_{+r}(r) \sigma^+_{(\mu} \sigma^r_{\nu)} + h_{+t}(r) \sigma^+_{(\mu} \sigma^t_{\nu)} + \\ 
        & h_{+3}(r) \sigma^+_{(\mu} \sigma^3_{\nu)} ) D_{\mathcal{J}\mathcal{M}}^\mathcal{J}) \,  ,
    \end{aligned} 
\end{equation}
\begin{equation}\label{eq:pertsimplyglobalscalar}
    \begin{aligned}
        h^S_{\mu\nu} & \equiv  e^{-i\omega \tau} r^2 ( h_{++}(r) \sigma^+_{\mu} \sigma^+_{\nu} D_{(\mathcal{J}-2)\mathcal{M}}^\mathcal{J} + \\ 
        & 2 (h_{+r}(r) \sigma^+_{(\mu} \sigma^r_{\nu)} + h_{+t}(r) \sigma^+_{(\mu} \sigma^t_{\nu)} + \\ 
        & h_{+3}(r) \sigma^+_{(\mu} \sigma^3_{\nu)} D_{(\mathcal{J}-1)\mathcal{M}}^\mathcal{J} ) + (2 h_{+-}(r) \sigma^+_{(\mu} \sigma^-_{\nu)} + \\ 
        & \sum_{i,j\in \{t,3,r\}} h_{ij}(r) \sigma^i_{(\mu} \sigma^j_{\nu)} ) D_{\mathcal{J}\mathcal{M}}^\mathcal{J}) \, ,
    \end{aligned}
\end{equation}
where we chose a convenient frame basis $\sigma^\pm = \frac{1}{2} \left(\sigma^1 \mp i \sigma^2 \right)$, indicated by the frame index $+$. 
One may write the Wigner-D functions as
\begin{equation}\label{eq:wigner_to_little_d}
    D^\mathcal{J}_{\mathcal{KM}}(\theta,\phi,\psi)=e^{-i(\mathcal{M}\phi+\mathcal{K}\psi)} d^\mathcal{J}_{\mathcal{KM}} (\theta) \, ,
\end{equation}
revealing the two Fourier-like exponential factors associating $\mathcal{M}$ with $\phi$ and $\mathcal{K}$ with $\psi$. 

Each sector can be considered separately, which is reminiscent of the standard procedure considering {e.g.}~sound modes separately from shear diffusion modes in the non-rotating fluid. Due to time-translation invariance the time-derivative can be replaced by its eigenvalue, the frequency through a Fourier transformation $h_{\mu\nu}(\tau)\propto e^{-i\omega \tau} h_{\mu\nu}(\omega)$: $\partial_\tau h_{\mu\nu}= -i \omega h_{\mu\nu}$, as usual. Similarly, the spatial partial derivatives can be replaced by the eigenvalues of the Wigner-D functions as follows~\cite{Murata:2007gv,Murata:2008xr,Murata:2008yx}
\begin{equation} \label{eq:partialEVs}
\begin{aligned}
        \partial_+D^\mathcal{J}_{\mathcal{KM}}&=\sqrt{(\mathcal{J}+\mathcal{K})(\mathcal{J}-\mathcal{K}+1)}D^\mathcal{J}_{\mathcal{K}-1\ \mathcal{M}} \, ,\\  \partial_-D^\mathcal{J}_{\mathcal{KM}}&=-\sqrt{(\mathcal{J}-\mathcal{K})(\mathcal{J}+\mathcal{K}+1)}D^\mathcal{J}_{\mathcal{K}+1\ \mathcal{M}} \, ,\\ 
        \partial_3 D^\mathcal{J}_{\mathcal{KM}}&=-i\mathcal{K} D^\mathcal{J}_{\mathcal{KM}}  \, ,
\end{aligned}
\end{equation}
where $\partial_\pm={e_\pm}^\mu\partial_\mu$, $\partial_3={e_3}^\mu\partial_\mu$,  with orthonormal frame covectors, $\sigma^\pm = \frac{1}{2} \left(\sigma^1 \mp i \sigma^2 \right)$. So, for example, $\partial_\psi = \sigma^a_\psi \partial_a = \partial_3$ 
and, thus, $\partial_\psi D^\mathcal{J}_{\mathcal{KM}} = - i \mathcal{K} D^\mathcal{J}_{\mathcal{KM}}$. 

As a result of the steps above, the fluctuation equations in a given sector (sound, momentum diffusion, or tensor) depend only on $\mathcal {J}$, $r$, and $\omega$ (not on $\mathcal{M}$, $\mathcal K$ or on $\partial_\theta,\, \partial_\phi,\, \partial_\psi$). 
The fluctuation equations in the momentum diffusion sector and sound sector are each still complicated sets of coupled differential equations relating three or seven fields, respectively. 
For illustration, the set of shear diffusion equations is provided in the Supplemental Material, Eq.~\eqref{eq:vectorFlucs}. The large black hole limit will simplify them and give rise to a powerful boost symmetry.

\emph{Large black hole limit.}--- 
In the hydrodynamic regime temperature is large, hence, we consider large AdS black holes for which the temperature increases monotonically with the horizon radius $r_+$. As a side effect, these black holes are safe from instabilities~\cite{Garbiso:2020puw,Murata:2008xr,Murata:2008yx,Murata:2007gv}. We impose this limit on the level of the metric fluctuation equations scaling the frequency and the angular momentum of the fluctuation 
\begin{equation}\label{eq:largeBHLimitJw}
\omega\to 2\alpha \nu r_+/L \, , \quad
\mathcal{J}\to\alpha j r_+/L\, , \quad \alpha\to\infty \, ,
\end{equation} 
simultaneously with the black hole horizon radius and radial coordinate limit given by~\eqref{eq:largeBHLimitR}, 
keeping leading order terms in $\alpha$. 
This may be visualized as zooming in on a small patch of the $S^3$ located at a large radius. Locally, this patch appears non-compact to fluctuations with long wavelengths, i.e.,~small $\nu,\, j\ll 1$ in the limit~\eqref{eq:largeBHLimitJw}. 
In this limit, using standard properties~\cite{Tajima:2015owa}, also the third angle, $\theta$, in the Wigner-D function is associated with a combination of eigenvalues $\mathcal{P}(\mathcal{J,K,M})$, such that $D^\mathcal{J}_{\mathcal{KM}}(\theta,\phi,\psi)\propto \exp[{-i(\mathcal{M}\phi+\mathcal{K}\psi + \mathcal{P} \theta)}]$. 
This closely resembles the Fourier modes $\exp[i(k_x x +k_y y+k_z z)]$ in non-compact Minkowski space and signals the emergence of translation and boost invariance with the effective boundary geometry now being $\mathbb{R}^{3,1}\sim \mathbb{R}_t\times \mathbb{R}^3$.  

This 
leads to the decoupling of several fluctuations, as pointed out above. 
Remarkably, it gives rise to another substantial simplification. All fluctuation equations around the rotating black hole can now be transformed into their form in a black hole at rest (originally stated in~\cite{Policastro:2002se,Policastro:2002tn}) by the following boost transformation~\cite{Garbiso:2020puw}:
\begin{equation} \label{eq:criticalPointTrajectory}
     \mathfrak{q}^2 = \frac{\left(a \nu + j \right)^2}{1-a^2} \, , \qquad 
     \mathfrak{w}^2 = \frac{\left(\nu + a j \right)^2}{1-a^2}\, .
 \end{equation}
Specifically, the four coupled shear diffusion  
equations reduce to a single equation, namely the shear diffusion fluctuation equation in a non-rotating fluid~\cite{Kovtun:2005ev}, see the Supplemental Material. 
The sound sector fluctuation equations with~\eqref{eq:criticalPointTrajectory} reduce similarly to the single master field equation in a non-rotating fluid given in~\cite{Kovtun:2005ev}. 

In summary, the frequencies and momenta of hydrodynamic modes in our rotating fluid can be analytically computed from the known~\cite{Kovtun:2005ev} hydrodynamic modes in a fluid at rest. 
This analytic relation is the boost transformation \eqref{eq:criticalPointTrajectory}.

%
\emph{Critical points \& hydrodynamic convergence.}---
Recent insights gained from the analysis of {\it critical points of spectral curves} 
provide a means to determine the radius of convergence of the hydrodynamic series in the linear regime~\cite{Grozdanov:2019kge,Grozdanov:2019uhi,Withers:2018srf,Heller:2020uuy,Heller:2020hnq}. 
In hydrodynamics the spectral curve  arises from the  determinant of a system of hydrodynamic fluctuation equations that encode the hydrodynamic dispersion relations. 
Hence, the hydrodynamic spectral curves are inherently implicit functions of frequency and momentum {i.e.}~\ $ P(\mathfrak{w},\mathfrak{q}^2) = 0$.  
For example, $P(\mathfrak{w},\mathfrak{q}^2) = v_s^2 \mathfrak{w}^2-\mathfrak{q}^2=0$ encodes the sound dispersion relation $\mathfrak{w}=\pm v_s \mathfrak{q}$. 
Critical points of hydrodynamic spectral curves can be used to determine the radius of convergence of the hydrodynamic gradient expansion in complex momentum space~\cite{Grozdanov:2016tdf,Grozdanov:2017kyl}. 
A subset of critical points are branch points from the point of view of complex analysis and, thus, the dispersion relation of a given mode is not analytic at such critical points. 
Not all of the critical points of a given spectral curve are branch points or any other type of singularity~\cite{Heller:2020hnq}. Hence, in general the radius of convergence based on the nearest critical point is a lower bound for the actual radius of convergence~\footnote{The actual radius of convergence may be larger, if it is set by one of the critical points further away from the origin of complex momentum space.}.

The hydrodynamic expansion is performed around vanishing momentum and frequency, i.e., around the origin of complex momentum space, $\mathfrak{w}=0,\, \vec{\mathfrak{q}}=0$. 
With that view, it is clear that those branch points closest to that origin determine the radius of convergence of the expansion. 
More precisely, considering the complex $\vec{\mathfrak{q}}$-space, the distance of the magnitude-wise smallest critical momentum $|\vec{\mathfrak{q}}_c|$ from the origin is a lower bound on the radius of convergence, $R_c = |\vec{\mathfrak{q}}_c|$.  
Critical points are defined as those frequencies $\mathfrak{w}$ and momenta $\mathfrak{q}$ which satisfy the constraints
\begin{equation}\label{eq:criticalPointConditions}
  P(\mathfrak{w},\mathfrak{q})|_{(\mathfrak{w}_c,\mathfrak{q}_c)} = 0 \,,  \quad   
  \partial_\mathfrak{w} P(\mathfrak{w},\mathfrak{q})|_{(\mathfrak{w}_c,\mathfrak{q}_c)} = 0\, ,
\end{equation}
where $P$ is a complex-valued function of the complex-valued frequency and momentum $(\mathfrak{w},\mathfrak{q})\in \mathbb{C}^2$, and $(\mathfrak{w}_c,\mathfrak{q}_c)$ denotes a discrete set of critical points of the spectral curve defined by the implicit function $P(\mathfrak{w},\mathfrak{q})=0$. 
In~\eqref{eq:criticalPointConditions}, we used the fact that the large black hole limit, \eqref{eq:largeBHLimitR} and~\eqref{eq:largeBHLimitJw}, implies that the fluctuations depend only on one of the three possible momentum directions, which leads us to consider critical points in one momentum-direction only in the relations above. 
As an example, consider the analytically continued shear diffusion mode defined by the spectral curve
\begin{equation}
    P_{\text{shear}}(\mathfrak{w},\mathfrak{q}) = \mathfrak{w} + i \mathfrak{q}^2 D(\mathfrak{w,q}) + \mathcal{O}(\mathfrak{q}^4) \, , \, \quad \mathfrak{w,q}\in \mathbb{C} \, ,
\end{equation}
with the diffusion coefficient $D$, which in general is a function of the complex momentum. To leading order in derivatives, this encodes the familiar form of the shear mode dispersion relation $\mathfrak{w}(\mathfrak{q}^2) = - i D \mathfrak{q}^2 + \mathcal{O}(\mathfrak{q}^4)$, with all quantities analytically-continued to be complex-valued.  

For the actual rotating QGP formed in heavy-ion collisions, the hydrodynamic spectral curve $P$ is not known. Thus, we here consider the spectral curve of a rotating $\mathcal{N}=4$ SYM plasma, in the nontrivial time- and space-dependent equilibrium state shown in Fig.\ \ref{fig:rotational_fluid}, as a substitute for the spinning QGP. 
In that SYM plasma at rest ($a=0$) the critical points closest to the origin of complex momentum space were found numerically to be given by~\cite{Grozdanov:2019uhi}
\begin{align}\label{eq:criticalPointsAtRest}
 \mathfrak{w}_c & \approx \pm 1-i,\, \mathfrak{q}_c^2 \approx \pm 2 i \quad \text{(sound)} \, , \\ \nonumber
 \mathfrak{w}_c & \approx \pm 1.4436414 - 1.0692250 i ,\, \\ \nonumber  \mathfrak{q}_c^2 & \approx 1.8906469 \pm 1.1711505 i \quad \text{(shear diffusion)} \, .
\end{align}
These are calculated from the holographically defined spectral curve $P(\wn,\qn)=0$, where $P$ is the determinant of all possible metric fluctuations evaluated at the AdS-boundary~\cite{Grozdanov:2019kge}, see the Supplemental Material. 

\begin{figure}
    \begin{center}
    \includegraphics[width=0.4\textwidth]{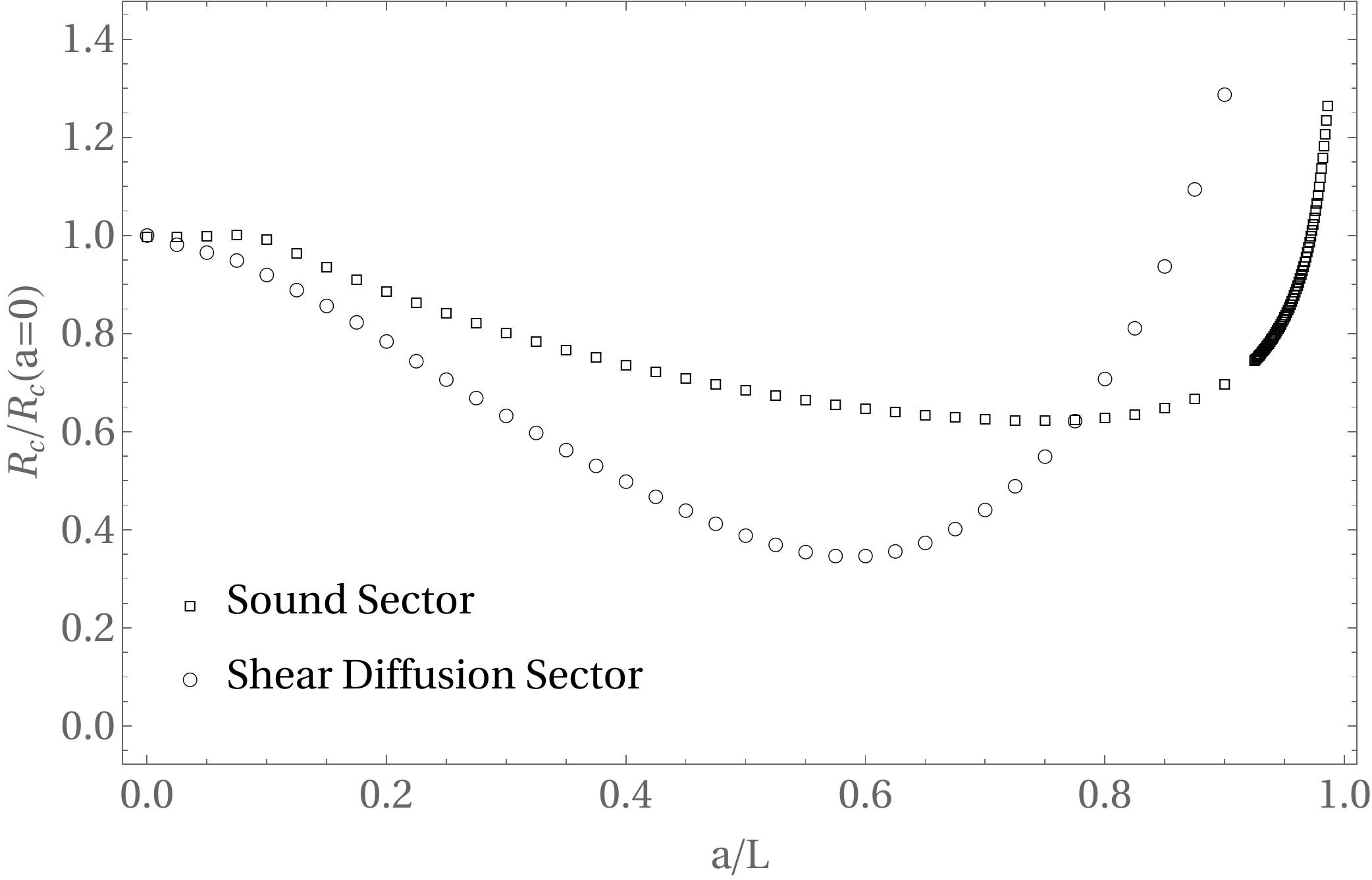}
    \end{center}
    \caption{The convergence radius, $R_c$, of the hydrodynamic expansion as a function of the angular momentum parameter $a/L$, normalized to its value in a fluid at rest ($a=0$). 
    }
    \label{fig:convergenceRadius}
\end{figure}

In the rotating case, the spectral curve can be expressed as $P(\nu,j)=0$ in terms of our frequency, $\nu$, and angular momentum variable, $j$. Here, $P$ is the determinant of the metric fluctuations~\eqref{eq:pertsimplyglobalscalar} and~\eqref{eq:pertsimplyglobalvector} evaluated at the AdS-boundary. This can be rewritten in terms of the non-rotating quantities as $P(\nu,j) =\bar{P}(\mathfrak{w},\mathfrak{q})$,  using~\eqref{eq:criticalPointTrajectory} in the invertible form 
${j} = \frac{ \mathfrak{q} - a \mathfrak{w}}{\sqrt{1-a^2}}$ and $\nu = \frac{\mathfrak{w}-a \mathfrak{q} }{\sqrt{1-a^2}}$.  
Thus, the critical point condition in the rotating plasma is given by 
\begin{subequations}
\begin{eqnarray}
    P(\nu,j)=0\,
    &&\Leftrightarrow \bar{P}(\wn,\qn)=0 \, ,\label{eq:criticalPointConditionsRotatingP} \\
     \partial_\nu P(\nu,j) =0\, &&\Leftrightarrow \partial_\wn \bar{P}(\wn,\qn)+a \partial_\qn \bar{P}(\wn,\qn)=0\, .
     \label{eq:criticalPointConditionsRotating}
\end{eqnarray} 
\end{subequations}
We have used the conditions on the non-rotating spectral curve $\bar{P}(\wn,\qn)$, given on the right side of~\eqref{eq:criticalPointConditionsRotatingP} and~\eqref{eq:criticalPointConditionsRotating}, in order to obtain Fig.~\ref{fig:convergenceRadius}. 
This method was verified by a second calculation in which we explicitly calculate numerically the closest critical point in our rotating system using the left side of~\eqref{eq:criticalPointConditionsRotatingP} and~\eqref{eq:criticalPointConditionsRotating} on the rotating spectral curve $P(\nu,j)=0$.

The resulting radius of convergence $R_c=\left|j_c\right|$ for each sector is given in Fig.~\ref{fig:convergenceRadius}. 
In the sound sector (squares),  the radius first increases at small angular momentum $a/L<0.075$, then it decreases to a minimum at $a/L\approx 0.75$. The lower bound on the radius of convergence in the sound sector is set by that minimum, and shows that the radius of convergence drops at worst to approximately 60\% of its value in a non-rotating fluid. 
In the shear diffusion sector (circles in Fig.~\ref{fig:convergenceRadius}) $R_c$ decreases monotonously to a minimum at $a/L\approx 0.6$.  
For large angular momentum beyond their respective minima,  $a>a_{\text{min}}$, the radius of convergence of the hydrodynamic expansion increases monotonously and quickly in both, sound and shear sector. 
In summary, the convergence is enhanced for all hydrodynamic modes in a rapidly rotating fluid with $a/L>0.95$, and the convergence radius takes a minimal value of approximately 60\% of its original value in the sound sector, 35\% in the shear sector. These are the lowest values the radius of convergence could be limited to due to rotation.  

We note that at vanishing angular momentum, $a=0$, the convergence radius of the sound mode, $R_c(a=0)\approx 1.41421$, differs slightly from the convergence radius of the shear diffusion mode, $R_c(a=0)\approx 1.49131$, in agreement with~\cite{Grozdanov:2019uhi}. 

\emph{Transport Coefficients.}--- 
At nonzero angular momentum, transport coefficients  split into the ones which measure transport along the angular momentum (longitudinal) and the ones measuring transport perpendicular to it (transverse). This is similar to the case of a charged plasma in the presence of a magnetic field, see \cite{Critelli:2014kra,Hernandez:2017mch,Ammon:2020rvg}. 
Transport coefficients in a fluid at rest have been extracted from the dispersion relations of the hydrodynamic QNMs~\cite{Kovtun:2005ev}. 
We compute the dispersion relations in a rotating fluid using the transformation~\eqref{eq:criticalPointTrajectory}. 
We also calculate the QNMs directly in the rotating fluid and confirm that the results agree. 
For example, when applying~\eqref{eq:criticalPointTrajectory} the shear diffusion mode dispersion relation in a fluid at rest given in~\cite{Kovtun:2005ev},  $\wn(\qn) = -i \qn^2/2 + \mathcal{O}(\qn^3)$, becomes
\begin{equation}
        \text{shear: } \nu(j) = -a j - i \frac{1}{2} (1-a^2)^{3/2} j^2 +\mathcal{O}(j^3) \,,
\end{equation}
where we now set $L=1$ for ease of notation for the remainder of this section. The sound diffusion dispersion relation is obtained in the same way 
\begin{equation}
         \text{sound: } \nu(j) =  \frac{\pm 1-\sqrt{3}a}{\sqrt{3}\mp a} j - i \sqrt{3}\frac{(1-a^2)^{3/2}}{(\sqrt{3}-a)^3} j^2 +\mathcal{O}(j^3) \, .
\end{equation}
From these computations, the following generalized relations are found for the longitudinal diffusion coefficient, $\mathcal{D}_{||}$, which turns into a damping of a mode now propagating with the speed of shear $v_{||}$, two modified speeds of sound $v_{s,\pm}$ and two sound attenuation coefficients $\Gamma_{\pm}$: 
\begin{eqnarray}
  v_{||} & = & a \, , \label{eq:vLSum}\\
  \mathcal{D}_{||} &=& \mathcal{D}_0 (1-a^2)^{3/2} \, ,\label{eq:DLSum4}\\
   v_{s,\pm} & = & v_{s,0} \frac{\sqrt{3} a\pm 1}{1\pm \frac{a}{\sqrt{3}}}\, ,\label{eq:vsSum}\\
 \Gamma_{s,\pm} & = & \Gamma_0 \frac{\left(1-a^2\right)^{3/2}}{\left(1\pm \frac{a}{\sqrt{3}}\right)^3} \, ,\label{eq:GsSum}
\end{eqnarray}
where for the holographic model the quantities at vanishing angular momentum, $a=0$, are:  $\mathcal{D}_0=1/2$, $v_{s,0}=1/\sqrt{3}$, and $\Gamma_0=1/3$.
Meanwhile, the two shear viscosities, $\eta_\perp$ and $\eta_{||}$ were computed in~\cite{Garbiso:2020puw} from the corresponding energy-momentum tensor two-point functions,  
according to the recipe~\cite{Policastro:2002se,Policastro:2002tn} 
\begin{equation}\label{eq:etas}
    \eta_{\perp}(a) = \eta_0 \frac{1}{\sqrt{1-a^2}}\, , \quad
  \eta_{||}(a) = \eta_0 {\sqrt{1-a^2}}\, ,
\end{equation}
and the shear viscosity at $a=0$ is given by $\eta_0=N^2 \pi T_0^3/8$. 
These quantities are related to each other by generalizations of their $a=0$ Einstein relations
\begin{eqnarray}\label{eq:DLSum}
 \mathcal{D}_{||}(a)
    &=&2 \pi T_0 \frac{\eta_{||}(a)}{\epsilon(a) + P_\perp(a)}\, , \nonumber \\
    \label{eq:GammaSum}
    \Gamma_\pm(a)&=&\frac{2\eta_{||}(a)}{3(\epsilon(a) + P_\perp(a))} \frac{1}{(1\pm a/\sqrt{3})^3} \, .
\end{eqnarray}
and the bulk viscosity $\zeta(a)=0$ in our conformal plasma.

\emph{Discussion.}---
We have computed two main results for rotating $\mathcal{N}=4$ SYM plasma: 
    (i) The radius of convergence, which quantifies the range of validity of the hydrodynamic gradient expansion, remains nonzero at nonzero angular momentum and increases for rapidly rotating fluids, see Fig.~\ref{fig:convergenceRadius}.  
    (ii) The values of the shear diffusion, shear viscosity, speed of shear propagation, speed of sound, sound attenuation in the rotating plasma are given analytically as a function of their values in the fluid at rest and as a function of the fluid angular momentum, see equations~\eqref{eq:vLSum}-\eqref{eq:DLSum}. 
    
    Our analysis reveals new aspects about the domain of applicability of hydrodynamics by determining the fate of the hydrodynamic series in a quantum system undergoing rapid rotation. This is the first time such a study has been performed at nonzero angular momentum, both in non-relativistic and relativistic systems. In fact, different than previous works which employed a constant and uniform equilibrium state in relativistic systems \cite{Grozdanov:2019uhi,Grozdanov:2019kge,Withers:2018srf,Abbasi:2020ykq,Jansen:2020hfd,Abbasi:2020xli}, here the convergence analysis is performed in a nontrivial time- and space-dependent flow which can display large spatial gradients induced by rotation. Assuming our results can be used as a proxy for the rapidly rotating QGP formed in heavy-ion collisions, our work indicates that hydrodynamics remains a good approximation to describe the evolution of this system even when it is subjected to gradients that can be already large enough to spoil the hydrodynamic expansion of its non-rotating counterpart. As such, these results provide useful guidance for future hydrodynamic simulations of heavy-ion collision with nonzero angular momentum. Finally, we remark that it would be a ground-breaking simplification if a boost symmetry such as~\eqref{eq:criticalPointTrajectory} could be derived for QCD based on the holographic blueprint we provide in this Letter.

\emph{Acknowledgments.}--- 
C.C., M.G.A., and M.K. were supported, in part, by the U.S.~Department of Energy grant DE-SC-0012447. J.N. is partially supported by the U.S. Department of Energy, Office of Science, Office for Nuclear Physics under Award No.\ DE-SC0021301. 
\bibliographystyle{apsrev4-1} 
\bibliography{bib}

\onecolumngrid

\appendix 

\section*{Supplemental Material}
This section demonstrates how the boost symmetry~\eqref{eq:criticalPointTrajectory} arises in the large black hole limit, leading to the significant simplification of the fluctuation equations discussed in the main text. In order to illustrate the vast complexity of the fluctuation equations, let us consider the example of the shear diffusion sector generically coupling four metric fluctuations, $h_{t+},\, h_{3+},\, h_{++}$ and $h_{r+}$. After choosing the standard radial gauge $h_{\mu r}\equiv 0$ (implying $h_{r+}\equiv 0$), there are three dynamical equations,
\begin{equation}\label{eq:vectorFlucs}
    \begin{aligned}
        0 = & h_{t+}''(r) + \frac{L^2 \left(2 a^2 \mu -10 \mu  r^2+5 r^4\right)+2 a^2 \mu  r^2+5 r^6}{L^2 \left(2 a^2 \mu  r-2 \mu  r^3+r^5\right)+2 a^2 \mu  r^3+r^7}h_{t+}'(r) + \\
        &\frac{8 a \mu  \left(L^2+2 r^2\right)}{L^2 \left(2 a^2 \mu  r-2 \mu  r^3+r^5\right)+2 a^2 \mu  r^3+r^7} h_{3+}'(r)+\\
        &\frac{L^2 h_{t+}(r)}{\left(L^2 \left(2 a^2 \mu  r-2 \mu  r^3+r^5\right)+2 a^2 \mu  r^3+r^7\right)^2} (-4 L^2 (4 a^4 \mu ^2-2 a^2 \mu ^2 r^2 (a \omega -2 \mathcal{J})\\
        &+\mathcal{J} \mu  r^6 (a \omega -2 \mathcal{J}-4)+\mathcal{J} (\mathcal{J}+2) r^8)-16 a^4 \mu ^2 r^2-4 \mathcal{J} (\mathcal{J}+2) r^{10})-\\
        &\frac{2 i \sqrt{2} \sqrt{\mathcal{J}} L^2 r^2 \omega }{L^2 \left(2 a^2 \mu -2 \mu  r^2+r^4\right)+2 a^2 \mu  r^2+r^6} h_{++}(r)-\\
        &\frac{4 L^2 h_{3+}(r)}{\left(L^2 \left(2 a^2 \mu  r-2 \mu  r^3+r^5\right)+2 a^2 \mu  r^3+r^7\right)^2} (-8 a^3 \mu ^2 \left(L^2+r^2\right)-2 a^2 \mu  r^2 \omega  \left(L^2 \left(2 \mu +r^2\right)+r^4\right)+\\
        &a \mu  r^2 \left(L^2 r^4 \omega ^2-4 (\mathcal{J}+2) \left(L^2 \left(r^2-2 \mu \right)+r^4\right)\right)+\mathcal{J} r^6 \omega  \left(L^2 \left(r^2-2 \mu \right)+r^4\right))\text{,}\\
        0 = & h_{++}''(r) + \frac{L^2 \left(3 r^4-2 \mu  \left(a^2+r^2\right)\right)+2 a^2 \mu  r^2+5 r^6}{L^2 \left(2 a^2 \mu  r-2 \mu  r^3+r^5\right)+2 a^2 \mu  r^3+r^7} h_{++}'(r) -\\
        &\frac{2 i \sqrt{2} \sqrt{\mathcal{J}} L^4 r^4 \left(2 a \mu  (a \omega -2 \mathcal{J}-2)+r^4 \omega \right)}{\left(L^2 \left(2 a^2 \mu -2 \mu  r^2+r^4\right)+2 a^2 \mu  r^2+r^6\right)^2} h_{t+}(r)+\\
        & \frac{h_{++}(r)}{\left(L^2 \left(2 a^2 \mu -2 \mu  r^2+r^4\right)+2 a^2 \mu  r^2+r^6\right)^2}(-4 (\mathcal{J}+1) L^2 r^2 (\mathcal{J} r^4 \left(L^2+r^2\right)-\\
        &2 \mu  \left(a^2 \left(L^2+r^2\right)+\mathcal{J} L^2 r^2\right))+L^4 r^4 \omega ^2 \left(2 a^2 \mu +r^4\right)-8 a (\mathcal{J}+1) \mu  L^4 r^4 \omega ) +\\
        & \frac{8 i \sqrt{2} \sqrt{\mathcal{J}} L^2 r^4 \left(a \mu  L^2 \omega +(\mathcal{J}+1) \left(L^2 \left(r^2-2 \mu \right)+r^4\right)\right)}{\left(L^2 \left(2 a^2 \mu -2 \mu  r^2+r^4\right)+2 a^2 \mu  r^2+r^6\right)^2} h_{3+} \text{,}\\
        0 = & h_{3+}''(r) -\frac{4 a \mu  L^2 r}{L^2 \left(2 a^2 \mu -2 \mu  r^2+r^4\right)+2 a^2 \mu  r^2+r^6} h_{t+}'(r) +\\
        & \frac{L^2 \left(6 a^2 \mu -2 \mu  r^2+3 r^4\right)+5 r^2 \left(2 a^2 \mu +r^4\right)}{L^2 \left(2 a^2 \mu  r-2 \mu  r^3+r^5\right)+2 a^2 \mu  r^3+r^7} h_{3+}'(r) -\\
        &\frac{L^4 \left(\mathcal{J} r^4-2 a^2 \mu \right) \left(2 a \mu  (a \omega -2 \mathcal{J}-2)+r^4 \omega \right)}{\left(L^2 \left(2 a^2 \mu -2 \mu  r^2+r^4\right)+2 a^2 \mu  r^2+r^6\right)^2} h_{t+}(r)-\\
        & \frac{2 i \sqrt{2} \sqrt{\mathcal{J}} (\mathcal{J}+1) L^2 r^2}{L^2 \left(2 a^2 \mu -2 \mu  r^2+r^4\right)+2 a^2 \mu  r^2+r^6} h_{++}(r)+\\
        & \frac{h_{+3}(r)}{\left(L^2 \left(2 a^2 \mu  r-2 \mu  r^3+r^5\right)+2 a^2 \mu  r^3+r^7\right)^2} (L^4 (-32 a^4 \mu ^2-8 a^2 \mu ^2 r^2 (a \omega -2 \mathcal{J}-6)-\\
        &16 a^2 (\mathcal{J}+2) \mu  r^4+2 \mu  r^6 (-2 \mathcal{J} (a \omega -4)+a \omega  (a \omega -4)+8)\\
        &-8 (\mathcal{J}+1) r^8+r^{10} \omega ^2)-8 L^2 r^2 \left(2 a^2 \mu +r^4\right) \left(2 a^2 \mu +(\mathcal{J}+1) r^4\right))  \text{,}
        \end{aligned}
 \end{equation}    
and one first order constraint equation,
       \begin{equation}
        \begin{aligned}
        0 = & h_{3+}'(r)-\frac{i \sqrt{\mathcal{J}} \left(L^2 \left(2 a^2 \mu -2 \mu  r^2+r^4\right)+2 a^2 \mu  r^2+r^6\right)}{\sqrt{2} \left(L^2 \left(-2 a^2 \mu +\mu  r^2 (a \omega -2 \mathcal{J})+\mathcal{J} r^4\right)-2 a^2 \mu  r^2+\mathcal{J} r^6\right)} h_{++}'(r)-\\
        &\frac{L^2 r^2 \left(2 a \mu  (a \omega -2 \mathcal{J}-2)+r^4 \omega \right)}{4 \left(L^2 \left(-2 a^2 \mu +\mu  r^2 (a \omega -2 \mathcal{J})+\mathcal{J} r^4\right)-2 a^2 \mu  r^2+\mathcal{J} r^6\right)} h_{t+}'(r) + \\
        &\frac{4 a \mu  L^2 r}{L^2 \left(-2 a^2 \mu +\mu  r^2 (a \omega -2 \mathcal{J})+\mathcal{J} r^4\right)-2 a^2 \mu  r^2+\mathcal{J} r^6} h_{t+}(r)-\\
        &\frac{8 a^2 \mu  \left(L^2+r^2\right)}{L^2 \left(-2 a^2 \mu  r+\mu  r^3 (a \omega -2 \mathcal{J})+\mathcal{J} r^5\right)-2 a^2 \mu  r^3+\mathcal{J} r^7} h_{3+}(r) \, .
    \end{aligned}
    \end{equation}
The $h_{++}$-component of the metric fluctuations decouples from this sector in the large black hole limit, Eqs.~\eqref{eq:largeBHLimitJw} and~\eqref{eq:largeBHLimitR}. 
This can be seen by considering the powers of $\alpha$ associated with each term in~\eqref{eq:vectorFlucs}. Consider, for example, the fifth line, where $h_{++}$ multiplies 
\begin{equation}
\frac{2 i \sqrt{2} \sqrt{\mathcal{J}} L^2 r^2 \omega }{L^2 \left(2 a^2 \mu -2 \mu  r^2+r^4\right)+2 a^2 \mu  r^2+r^6} \, ,
\end{equation}
which is of order $\sqrt{\mathcal{J}} \omega/r^4 \sim \alpha^{3/2}/\alpha^4 \sim \alpha^{-5/2}$ and, therefore, is suppressed compared to the leading order terms which are of order $\alpha^{-2}$ as the term including $h_{3+}$ in the sixth line exemplifies. 
This decoupling can be understood to occur due to a symmetry enhancement stemming from the limit~\eqref{eq:largeBHLimitJw}. 
This leaves the two coupled fields $h_{t+},\, h_{3+}$, and their equations of motion to have a similar structure as those for $h_{tx},\, h_{zx}$ in the non-rotating case~\cite{Kovtun:2005ev}, written in Poincar{\'e} coordinates $(t,x,y,z,r)$. 
In fact, as pointed out in the main text, the transformation~\eqref{eq:criticalPointTrajectory} identically maps the two sets of shear diffusion equations into each other. 
In a similar way,  $h_{++},\, h_{+3},\, h_{+t}$ decouple from the sound sector. The (excessively long) sound sector equations are available in form of a Mathematica notebook upon request. 

When we apply the large black hole limit~\eqref{eq:largeBHLimitJw} and \eqref{eq:largeBHLimitR} to the vector fluctuation equations~\eqref{eq:vectorFlucs}, and use the boost-like transformation~\eqref{eq:criticalPointTrajectory} from $(\nu,j)$ to $(\mathfrak{w,q})$, we obtain the master field equation
\begin{equation}\label{eq:vectorFlucNonRotating}
 Z_1'' +  \frac{(\wn^2 - \qn^2 f)f - u\wn^2 f'}{uf(\qn^2 f -\wn^2)} Z_1'
    + \frac{\wn^2 - \qn^2 f}{u f^2} Z_1  =  0 \, , 
\end{equation}
with the inverted radial coordinate $u=r_+^2/r^2$, blackening factor $f=1-u^2$, and written in terms of the master field 
$Z_1(u)\equiv \qn u h_{tx}/(\pi T L)^2 + \wn u h_{zx}/(\pi T L)^2$. 
This latter equation is that for shear diffusion modes in a fluid at rest~\cite{Kovtun:2005ev}. 
This master field evaluated at the AdS-boundary, $u=0$, defines the holographic spectral curve $P(\wn,\qn)=Z_1(\wn,\qn;u=0)=0$~\cite{Grozdanov:2019kge}.

\bigskip

Note that the simplifications in the large black hole limit may alternatively be understood based on the eigenvalue equations~\eqref{eq:partialEVs}. In the hydrodynamic limit~\eqref{eq:largeBHLimitR} and~\eqref{eq:largeBHLimitJw}, only leading terms of order $\alpha^{2}$ survive, which can stem from second derivatives only: $\partial_t^2\sim \omega^2\sim \alpha^2$, $\mathcal{J}^2\sim \alpha^2$, or $\omega \mathcal{J}\sim \alpha^2$. 
Furthermore, ~\eqref{eq:largeBHLimitR} and~\eqref{eq:largeBHLimitJw} imply $\mathcal{J,K}\to \infty$, which leads to $\mathcal{K}+n \approx\mathcal{J}$ for $n=0,\pm 1,\pm 2$. Then, \eqref{eq:partialEVs} becomes 
\begin{equation} \label{eq:partialEVsLargeBH}
\begin{aligned}
        \partial_+D^\mathcal{J}_{\mathcal{JM}}& \approx\sqrt{2\mathcal{J}}D^\mathcal{J}_{\mathcal{J}\ \mathcal{M}} \, ,\\  \partial_-D^\mathcal{J}_{\mathcal{JM}}& \approx 0 \, ,\\ 
        \partial_3 D^\mathcal{J}_{\mathcal{JM}}&\approx-i\mathcal{J} D^\mathcal{J}_{\mathcal{JM}}  \, .
\end{aligned}
\end{equation}
This indicates that $\partial_+$ can at most contribute $\partial_+^2\sim \mathcal{J}\sim\alpha$, while $\partial_-$ does not contribute to the fluctuation equations. 
Hence, all factors of the continuous momentum $j$ surviving the limit~\eqref{eq:largeBHLimitR} and~\eqref{eq:largeBHLimitJw} stem from $\partial_3$-derivatives acting on Wigner-D functions.

\end{document}